\documentstyle[aps,epsf,preprint]{revtex}
\draft

\begin{document}



\newcommand{\refkl}[1]{(\ref{#1})}
\newcommand{\tabsetting}{
              \renewcommand{\baselinestretch}{1.0}\small \normalsize}


\newcommand{\Angstroem}{{\AA}}
\newcommand{\cels}{\mbox{$^{\circ}{\rm C}$}}
\newcommand{\Fr}{\mbox{Fr\'{e}edericksz--}}
\newcommand{\Poincare}{Poincar\'{e}}
\newcommand{\rb}{Rayleigh--B\'{e}nard}
\newcommand{\RB}{Rayleigh--B\'{e}nard}
\newcommand{\via}{{\it via}}


\newcommand{\tabcondexp}{I}
\newcommand{\tabcondthick}{II}
\newcommand{\tabmatpar}{III}

\newcommand{\titlebox}[1]{\parbox{140mm}{\vspace{2mm} #1 \vspace{2mm}}}

\newcommand{\spacelinebot}[1]{\parbox[t]{1mm}{\hspace{1mm}\vspace{#1}}}
\newcommand{\spacelinetop}[1]{\parbox[b]{1mm}{\hspace{1mm}\vspace{#1}}}
\newcommand{\spacelinemid}[1]{\parbox[c]{1mm}{\hspace{1mm}\vspace{#1}}}
\newcommand{\spacingbot}{\parbox[t]{1mm}{\hspace{1mm}\vspace{3mm}}}
\newcommand{\spacingtop}{\parbox[b]{1mm}{\hspace{1mm}\vspace{6mm}}}
\newcommand{\spacingmid}{\parbox[c]{1mm}{\hspace{1mm}\vspace{9mm}}}


\newcommand{\oldfigc}[2]{
   \renewcommand{\baselinestretch}{1.0}
   \parbox[t]{#1}{\sloppy \small #2}
   \renewcommand{\baselinestretch}{\usualstretch}
   \small \normalsize
   }


\newcommand{\figps}[2]{
   \begin{minipage}[]{#1}
       \epsfxsize #1
       \epsffile{#2}
   \end{minipage}
   }

\newcommand{\figcaption}[3]{
   \renewcommand{\baselinestretch}{1.0}
   \noindent
   \parbox[]{#1}
      {\sloppy \small Figure #2 \
       #3
       }
    \renewcommand{\baselinestretch}{\usualstretch}
    \small \normalsize
   }


\newcommand{\largefig}[4]{    
   \begin{minipage}{\lenxtot}
       \figps{\lenxtot}{#1}
       \vspace{#2} \\
       \figcaption{\lenxtot}{#3}{#4}
   \end{minipage}
   }


\newcommand{\smallfig}[5]{    
   \begin{minipage}{\lenxtot}
       \begin{minipage}{#1}
         \figps{#1}{#2}
       \end{minipage}
       \hfill
       \begin{minipage}{#3}
         \figcaption{#3}{#4}{#5}
       \end{minipage}
   \end{minipage}
   }

\newcommand{\abl}[2]{\frac{\partial #1}{\partial #2}}  
\newcommand{\abltot}[2]{\frac{d #1}{d #2}}  
\newcommand{\ablzwei}[2]{\frac{\partial^{2} #1}{\partial #2^{2}}}
\newcommand{\ablii}[2]{\frac{\partial^{2} #1}{\partial #2^{2}}}
\newcommand{\ablgem}[3]{\frac{\partial^{2} #1}{\partial #2\partial #3}} 
\newcommand{\boxsub}[1]{_{\mbox{{\scriptsize #1}}}}
\newcommand{\boxsup}[1]{^{\mbox{{\scriptsize #1}}}}
\newcommand{\cc}{^{\ast}}         
\renewcommand{\d}[1]{\partial_{#1}}                     
\newcommand{\deltat}{\mbox{$\delta(t-t')$}}
\newcommand{\deltar}{\mbox{$\delta(\v{r}-\v{r'})$}}
\newcommand{\deltax}{\mbox{$\delta(x-x')$}}
\newcommand{\deltavx}{\mbox{$\delta(\v{x}-\v{x}')$}}
\newcommand{\deltay}{\mbox{$\delta(y-y')$}}
\newcommand{\deltaz}{\mbox{$\delta(z-z')$}}
\newcommand{\deltart}{\mbox{$\delta(\v{r}-\v{r'})\delta(t-t')$}}
\newcommand{\deltaxyt}{\mbox{$\delta(x-x')\delta(y-y')\delta(t-t')$}}
\newcommand{\deltaij}{\mbox{$\delta_{ij}$}}
\newcommand{\deltaik}{\mbox{$\delta_{ik}$}}
\newcommand{\deltail}{\mbox{$\delta_{il}$}}
\newcommand{\deltajk}{\mbox{$\delta_{jk}$}}
\newcommand{\deltajl}{\mbox{$\delta_{jl}$}}
\newcommand{\erw}[1]{\mbox{$<\!#1\!>$}} 
\newcommand{\funkint}[1]{\int \!\!\cal{D}[#1]}
\newcommand{\hc}{^{\dagger}}         
\renewcommand{\Im}{\mbox{Im}}
\newcommand{\intd}[1]{\int \!\!d#1}
\newcommand{\intdn}[1]{\int \!\!d^{n}\v{#1}}
\newcommand{\intvol}{\int \!\!d^{3}} 
\newcommand{\intvolr}{\int \!\!d^{3} r} 
\newcommand{\m}[1]{\underline{\underline{#1}}}           
\newcommand{\nab}{\v{\nabla}}
\newcommand{\order}[1]{{\cal O}(#1)}
\newcommand{\overdot}[1]{\stackrel{.}{#1}}
\renewcommand{\Re}{\mbox{Re}}
\newcommand{\re}{\mbox{Re}}
\newcommand{\rot}[1]{\v{\nabla}\times \v{#1}}
\renewcommand{\v}[1]{\mbox{\boldmath$#1$}}     
\newcommand{\vscript}[1]{\mbox{{\scriptsize $\bf #1$}}} 
\newcommand{\varabl}[2]{\frac{\delta #1}{\delta #2}}  


\newcommand{\vonx} {\mbox{$(\v{x})$}}                   
\newcommand{\vonk} {\mbox{$(\v{k})$}}                   
\newcommand{\vonw} {\mbox{$(\omega )$}}                   
\newcommand{\vonr} {\mbox{$(\v{r})$}}                   
\newcommand{\vonrs} {\mbox{$(\v{r'})$}}                   
\newcommand{\vonxyt} {\mbox{$(x,y,t)$}}                   
\newcommand{\vongrad} {\mbox{$(\v{\nabla })$}}            
\newcommand{\vonrgrad} {\mbox{$(\v{r},\v{\nabla })$}}     
\newcommand{\vonxt} {\mbox{$(\v{x},t)$}}                
\newcommand{\vonrt} {\mbox{$(\v{r},t)$}}                
\newcommand{\vonrsts} {\mbox{$(\v{r'},t')$}}                

\newcommand{\nonu} {\nonumber}



\newcommand{\abseps}{|\epsilon|}
\newcommand{\abslamsig}{|\lambda_{\sigma}|}
\newcommand{\aiiabs}{|\alpha_2|}
\newcommand{\Ap}{\mbox{A$^+$}}
\newcommand{\aschl}{\tilde{\alpha}}
\newcommand{\Bm}{\mbox{B$^-$}}
\newcommand{\barsig}{\overline{\sigma}}
\newcommand{\barphi}{\overline{\phi}}
\newcommand{\barrho}{\overline{\rho}}
\newcommand{\barn}{\overline{n}}
\newcommand{\baru}{\overline{u}}
\newcommand{\barv}{\overline{v}}
\newcommand{\crit} {|_c}
\newcommand{\cz} {\sqrt{2/\pi}\cos z}
\newcommand{\deltaperp}{\m{\delta}^{\perp}}
\newcommand{\dijp}{\delta_{ij}^{\perp}}
\newcommand{\dikp}{\delta_{ik}^{\perp}}
\newcommand{\drs} {\partial_r^{\ast}}
\newcommand{\dEo}{\delta E_0}
\newcommand{\deltaehc}{\delta_{EHC}}
\newcommand{\deltarb}{\delta_{RBC}}
\newcommand{\Deltaq}{\Delta_q}
\newcommand{\Dg}{d(\gamma)}
\newcommand{\Drho}{D_{\rho}}
\newcommand{\drhoo}{\delta \rho_0}
\newcommand{\Dphys}{D^{phys}}
\newcommand{\Ds}{D_{\sigma}}
\newcommand{\dschl}{\tilde{d}}
\newcommand{\Dschl}{\tilde{D}}
\newcommand{\dsigo}{\delta\sigma_0}
\newcommand{\ea}{\epsilon_a}
\newcommand{\eff}{^{\mbox{{\scriptsize (eff) }}}}
\newcommand{\Eeff}{\overline{E}}
\newcommand{\Erms}{\overline{E}}
\newcommand{\ehc} {^{(E)}}
\newcommand{\Esrb}{\erw{E'}\subrbc}
\newcommand{\Estc}{\erw{E'}\subtcf}
\newcommand{\Esehc}{\erw{E'}\subehc}
\newcommand{\epsa}{\epsilon_{a}}
\newcommand{\eaeff}{\epsilon_a\eff}
\newcommand{\eo}{\epsilon_0}
\newcommand{\ep}{\epsilon_{\perp}}
\newcommand{\eps}{\epsilon}
\newcommand{\epsq}{\epsilon_{q}}
\newcommand{\epsp}{\epsilon_{\perp}}
\newcommand{\epsperp}{\epsilon_{\perp}}
\newcommand{\epspar}{\epsilon_{\parallel}}
\newcommand{\epso}{\epsilon_0}
\newcommand{\eq}{\boxsup{eq}}
\newcommand{\Erho}{E_{\rho}}
\newcommand{\Es}{E_{\sigma}}
\newcommand{\etaeff}{\eta\eff}
\newcommand{\fpR} {\v{f}^{\dagger R}}
\newcommand{\gammasig}{\gamma_{\sigma}}
\newcommand{\gsig}{g_{\sigma}}
\newcommand{\geomu}{\sqrt{\mupp\mupm}}
\newcommand{\gSM}{g\SM}
\newcommand{\hcrb} {^{\dagger(RBC)}}
\newcommand{\hccc} {^{\dagger\ast}}
\newcommand{\hochi} {^{(1)}}
\newcommand{\hochii} {^{(2)}}
\newcommand{\hochiii} {^{(3)}}
\newcommand{\hocho} {^{(0)}}
\newcommand{\hochv} {^{(v)}}
\newcommand{\hochn} {^{(n)}}
\newcommand{\hochel} {^{(el)}}
\newcommand{\hochT} {^{(th)}}
\newcommand{\hochs} {^{(s)}}
\newcommand{\hochas} {^{(as)}}
\newcommand{\hot} {\mbox{h.o.t.}}
\newcommand{\keff}{K\eff}
\newcommand{\Ivii}{\mbox{I 52}}
\newcommand{\intpmpidz}{\int^{\pi/2}_{\pi/2}\! dz}
\newcommand{\Jeps}{\v{J}_{\eps}}
\newcommand{\JepsD}{\v{J}_{\eps}^D}
\newcommand{\Jnp}{\v{J}_{n^{+}}}
\newcommand{\Jnpm}{\v{J}_{n^{\pm}}}
\newcommand{\Js}{\v{J}_s}
\newcommand{\JsD}{\v{J}_s^D}
\newcommand{\jump}{\boxsub{jp}}
\newcommand{\lambdaeff}{\lambda\eff}
\newcommand{\lamBL}{\lambda\boxsub{BL}}
\newcommand{\lamD}{\lambda_{D}}
\newcommand{\lamsig}{\lambda_{\sigma}}
\newcommand{\lamBLphys}{\lambda\boxsub{BL}/boxsup{phys}}
\newcommand{\Lewis}{{\cal L}}
\newcommand{\lif}{\boxsub{lif}}
\newcommand{\miiVs}{\mbox{m}^2/(\mbox{Vs})}
\newcommand{\mup}{\mu^{+}}
\newcommand{\muperp}{\mu_{\perp}}
\newcommand{\mupp}{\mu_{\perp}^{+}}
\newcommand{\mum}{\mbox{$\mu \mbox{m}$}}
\newcommand{\mupm}{\mu_{\perp}^{-}}
\newcommand{\muppm}{\mu_{\perp}^{\pm}}
\newcommand{\mmupm}{\m{\mu}^{\pm}}
\newcommand{\nabperp} {\nabla_{\perp}}
\newcommand{\nabpar} {\nabla_{\parallel}}
\newcommand{\nhocho}{n^{(0)}}
\newcommand{\NL}{\boxsub{NL}}
\newcommand{\no}{n^{(0)}}
\newcommand{\np}{n^{+}}
\newcommand{\nm}{n^{-}}
\newcommand{\npm}{n^{\pm}}
\newcommand{\nzbar}{\overline{n}_z}
\newcommand{\nzqc}{\overline{n}_{z,q_c}}
\newcommand{\nbar}{\overline{n}}
\newcommand{\Ommi}{(\Omega \mbox{m})^{-1}}
\newcommand{\omr} {\omega(\overline{r})}
\newcommand{\oplamsig}{\hat{\lambda}_{\sigma}}
\newcommand{\opsigaeff}{\hat{\sigma}_a\eff}
\newcommand{\oplam}{\hat{\lambda}}
\newcommand{\opsig}{\hat{\sigma}}
\newcommand{\opeps}{\hat{\epsilon}}
\newcommand{\opsigq}{\hat{\sigma}_q}
\newcommand{\opepsq}{\hat{\epsilon}_q}
\newcommand{\opK}{\hat{K}}
\newcommand{\opKq}{\hat{K}_q}
\newcommand{\osc}{\boxsub{osc}}
\newcommand{\phihochi}{\phi^{(1)}}
\newcommand{\phihochmi}{\phi^{(-1)}}
\newcommand{\phip}{\phi^{+}}
\newcommand{\phim}{\phi^{-}}
\newcommand{\phys}{^{\mbox{\scriptsize phys}}}
\newcommand{\qcdotx}{\vscript{q}\cdot\vscript{x}}
\newcommand{\qccdotx}{\vscript{q}_c\cdot\vscript{x}}
\newcommand{\rec}{\boxsub{rec}}
\newcommand{\rhop}{\rho^{+}}
\newcommand{\rhom}{\rho^{-}}
\newcommand{\rhooschl}{\tilde{\rho}_0}
\newcommand{\rosm}{R_0\SM}
\newcommand{\Rosm}{R_0\SM}
\newcommand{\rbar}{\overline{r}}
\newcommand{\rperp} {r_{\perp}}
\newcommand{\rpar} {r_{\parallel}}
\newcommand{\rschl}{\tilde{r}}
\newcommand{\rtri}{\tilde{r}_{tri}}
\newcommand{\Sgi}{s_1(\gamma)}
\newcommand{\Sgii}{s_2(\gamma)}
\newcommand{\SM}{^{\mbox{{\tiny SM}}}}
\newcommand{\sig}{\sigma}
\newcommand{\siga}{\sigma_{a}}
\newcommand{\sigaeff}{\sigma_a\eff}
\newcommand{\sigbar}{\overline{\sigma}}
\newcommand{\sigq}{\sigma_{q}}
\newcommand{\sigo}{\sigma_0}
\newcommand{\sigp}{\sigma_{\perp}}
\newcommand{\sigpeq}{\sigma_{\perp}^{eq}}
\newcommand{\sigpar}{\sigma_{\parallel}}
\newcommand{\sigperp}{\sigma_{\perp}}
\newcommand{\sigtot}{\sigma_0}
\newcommand{\subrbc} {_{RBC}}
\newcommand{\subehc} {_{EHC}}
\newcommand{\subtcf} {_{TCF}}
\newcommand{\sz} {\sqrt{2/\pi}\sin z}
\newcommand{\siiz} {\sqrt{2/\pi}\sin 2z}
\newcommand{\stat} {^{\mbox{{\scriptsize stat}}}}
\newcommand{\talpha}{\tilde{\alpha}}
\newcommand{\tcf} {^{(T)}}
\newcommand{\rbc} {^{(R)}}
\newcommand{\tsii}{\tilde{\sigma}_{11}}
\newcommand{\tsigii}{\tilde{\sigma}_{11}}
\newcommand{\tr}{\tilde{r}}
\newcommand{\taudo}{\tau_d^{(0)}}
\newcommand{\tildesig}{\tilde{\sigma}}
\newcommand{\tildephi}{\tilde{\phi}}
\newcommand{\tilderho}{\tilde{\rho}}
\newcommand{\tilden}{\tilde{n}}
\newcommand{\tildevecu}{\tilde{\v{u}}}
\newcommand{\tildevecv}{\tilde{\v{v}}}
\newcommand{\tosm}{\tau_0\SM}
\newcommand{\tri}{\boxsub{tri}}
\newcommand{\ubar}{\overline{u}}
\newcommand{\uvec}{\v{u}}
\newcommand{\ui}{\v{u}^{(1)}}
\newcommand{\uqA}{\v{u}_q^A}
\newcommand{\uqB}{\v{u}_q^B}
\newcommand{\uqc}{\v{u}_{q_c}}
\newcommand{\umqc}{\v{u}_{-q_c}}
\newcommand{\us}{\v{u}_s}
\newcommand{\up}{\uparrow}
\newcommand{\down}{\downarrow}
\newcommand{\Vco}{V_{c0}}
\newcommand{\Vosm}{V_0^{SM}}
\newcommand{\Vrms}{\overline{V}}
\newcommand{\wH}{\omega_H}
\newcommand{\wh}{\omega_H}
\newcommand{\whphys}{\omega_H^{phys}}
\newcommand{\wschl}{\tilde{\omega}}
\newcommand{\ws}{\omega'}
\newcommand{\wo}{\omega_0}
\newcommand{\wophys}{\omega_0^{phys}}

\title{Travelling waves in electroconvection of the nematic Phase 5: \\
A test of the weak electrolyte model}
\author{Martin Treiber$^{\ast}$,
        N{\'a}ndor  {\'E}ber$^{\ast\ast}$,
        {\'A}gnes  Buka$^{\ast\ast}$,
        and Lorenz Kramer$^{\ast}$
       }
\address{$^{\ast}$ Universit{\"a}t Bayreuth, Theoretische Physik II,
         Universit{\"a}tsstra{\ss}e 30, D-95440 Bayreuth, Germany}
\address{$^{\ast\ast}$ Research Institute for Solid State
Physics of the Hungarian Academy of Sciences, H-1525 Budapest,
P.O.B.49, Hungary}

\date{\today}
\maketitle
%
%
\vspace{10mm}

\begin{center}
\parbox{140mm}{
We investigated
travelling waves  appearing as
the primary pattern-forming instability
in the nematic Phase 5 (Merck) in the planar geometry
in order to test the recently developed weak electrolyte model
of ac-driven electroconvection 
[M.~Treiber and L.~Kramer, Mol. Cryst. Liq. Cryst {\bf 261}, 311 (1995)].
Travelling waves are observed
over the full conductive range of applied frequencies
for three cells of different layer thickness $d$.
We also measured the elastic constants, the electric
conductivity, and the dielectric constant.
The other parameters of Phase 5 are known,
apart from the (relatively unimportant) viscosity $\alpha_1$
and the two parameters of the weak electrolyte model
that are proportional to the geometric mean
of the mobilities, and the recombination rate, respectively.
Assuming a sufficiently small recombination rate,
the predicted  dependence of the frequency of the travelling waves
at onset (Hopf frequency)
on $d$
and on the applied frequency
agreed quantitatively with the experiments, essentially
without fit parameters.
The absolute value of the Hopf frequency implies
that the geometric mean
of the mobilities amounts to $1.1 \times 10^{-10} \miiVs$. 
}
\end{center}

\newpage

%

\section{Introduction}
Electrohydrodynamic convection (EHC) in nematic liquid crystals (NLC) 
is one of the most prominent phenomena employed for the study
of pattern formation in anisotropic
systems\cite{kramer-EHCrevs,rehberg-sol,kramer-jphys}.
The system consists of a NLC
with negative or only mildly positive
dielectric anisotropy sandwiched between two
glass plates with transparent electrodes.
When applying an ac voltage with the circular frequency
$\wo$ and increasing the rms $\Vrms$ above a
certain threshold $\Vrms_c (\wo)$, the non-convecting basic state
becomes unstable to stationary or travelling rolls.
In this work, we consider the usual planar-homogeneous
geometry where the director of the NLC is aligned parallel to the plate
electrodes.

The traditionally used hydrodynamic model of EHC originating from
Helfrich \cite{helfrich} involves equations for the velocity field, 
the director, and the electric potential (or charge density). The NLC is
considered as an anisotropic ohmic conductor.
The origin of the conductivity is not specified.
This standard hydrodynamic description \cite{erickson,leslie,degennes},
including the three-dimensional formulation for EHC
\cite{zimmermann-oblique,kramer-jphys,kramer-EHC89}, is referred to
as the Standard Model (SM).
For a general introduction, see, e.g., the
book of Blinov \cite{blinov-old}, or the reviews
\cite{kramer-EHCrevs}.

Many features in the conductive
range of low ac frequencies are quantitatively described by the
SM \cite{kramer-jphys}. Examples are the threshold voltage
as a function of the ac frequency $\wo$,
the wave vector of the pattern, which describes the spacing and the angle
of the rolls and includes the possibility of oblique rolls
(where the roll axis is not perpendicular to the direction of planar  
alignment),
\cite{kramer-jphys,kramer-EHC89}
and, in the nonlinear regime, the amplitude of the roll pattern and  
secondary
bifurcations leading in particular to (weak) defect turbulence
\cite{kaiser-pesch-EHC93,decker-94,decker-PhD}.
Other features remain unexplained even on a qualitative level.
Most notable are travelling waves (TW) which have been observed
as early as 1978 \cite{kai}.
Later on, they were found by various groups in a broad parameter range in
the NLCs MBBA \cite{rehberg-PRL-89,rehberg-PRL-91-fluct},
I52
\cite{dennin-treiber-PRL,dennin-ILCC},
and in Phase 5
\cite{rehberg-sol,ribotta-88}.
TWs appear to be generic for relatively thin and clean cells
and seem to be often favoured by higher external frequencies.

The recently developed weak electrolyte model (WEM)
\cite{treiber-ILCC} predicts a Hopf bifurcation for these conditions,
provided the recombination rate of mobile ions
(see below) is sufficiently low.
The WEM equations
for the director and velocity fields are those of the SM, but the  
assumption
of ohmic conduction is dropped and the conduction properties
are explicitly modelled by
two species of oppositely charged
ionic charge carriers.
The dynamics of the number densities $\np\vonrt$ and $\nm\vonrt$
of the charge carriers is governed
by drift relative to the fluid  under an electric field
and by a conventional
dissociation-recombination reaction between neutral molecules and
the ions. The WEM expresses the
total space-charge density,
which already appears in the SM,
as the difference of the number densities of the two ionic species.
In addition, the drift of the charges excites
a new field, which can be
expressed in terms of a  variable local conductivity $\sigp \vonrt$
perpendicular to the director ("charge-carrier mode").


The SM has two relevant time scales: 
the director-relaxation time
$\tau_d = \gamma_1 d^2/(K_{11}\pi^2)$
and the charge-relaxation time
$\tau_q = \epsilon_{0} \epsilon_{\perp} /\sigma_{\perp}$.
Here $\epsilon_{\perp}$ is the principal
value of the dielectric-constant tensor perpendicular to the
director, $\gamma_1$ is a rotational viscosity,
and $K_{11}$ is the orientational elasticity for splay distortions.
Typically, $\tau_d$ is ${\cal O}(1\ {\rm s})$, and $\tau_q$ is
${\cal O}(10^{-2}\ {\rm s})$. 
The WEM has two additional time scales \cite{treiber-ILCC,dennin-treiber-PRL}:
a recombination time $\tau\rec$ for the relaxation towards the
equilibrium of the dissociation-recombination reaction where
$\sigp\vonrt = \sigp\eq$,
and the transition time $\tau_t = d^2 / (\Vrms \mu)$
for a charge with the mobility $\mu = \mupp + \mupm$ to traverse the
cell under the critical applied voltage.
The mobilities $\mupp$ and $\mupm$  
are the principal values
of the mobility tensors 
perpendicular to the director,
see Eq. \refkl{mu} below.
The magnitude of $\tau_t$ will turn out to be
of the order of 0.1 s whereas $\tau\rec > 5$ s.

The SM is recovered in the limits
$\tau_t/\tau_q >> 1$,
$\tau_t \omega_0 >> 1$,
and $\frac{\tau_t}{\tau\rec} \sqrt{\tau_d/\tau_q } >> 1$.
The ratio $\tau_t/\tau_q$ is a measure for the
relative change of the charge-carrier densities induced by the
space charges of the Carr--Helfrich mechanism.
In our case the third inequality will be violated.



%
%
%

Assuming a
recombination rate $\tau\rec^{-1}$ that is small compared to the
circular Hopf frequency $\wh = 2 \pi f_H$,
the WEM predicts the Hopf frequency $f_H$ to be proportional to
$(\mupp\mupm)^{1/2} (\sigp\eq)^{-1/2} d^{-3}$, and depending otherwise
only on $\wo$ and on the material parameters of the SM.
In experiments on the nematic I52 \cite{dennin-treiber-PRL},
the WEM explains succesfully  the dependence of
the Hopf frequency of the observed TWs on $\sigp\eq$ and on $\wo$.
Measurements on a thicker cell seem also to confirm the predicted
$d$ dependence, but no systematic measurements
were made.
Also, not all of the
twelve material parameters of the SM are known for I52. Six  
parameters had
to be fitted to the curves of the threshold voltage and of the
roll angle as a function of $\wo$.
These curves can be calculated, in a good approximation,
by the SM.

\medskip

In this work we present, for the nematic Phase 5 (Merck),
tests of the WEM essentially without fits of the SM parameters,
and we investigate systematically the $d$ dependence of
the Hopf frequency.

The threshold $\Vrms_c$, the wavevector $\v{q}_c$,
and the Hopf frequency
$\wh$ of the travelling rolls
are measured, as a function
of $\wo$, for three cells of different layer thickness $d$.
We also measured the orientational-elastic moduli
for splay $K_{11}$, twist $K_{22}$, and bend $K_{33}$,
the equilibrium conductivity $\sigp\eq$,
and the dielectric permittivity $\epsp$
perpendicular to the director as well as their anisotropies
($\siga\eq/\sigp\eq$ and $\ea$).
Then, together with known data for the viscosities
\cite{kneppe}, all SM parameters of Phase 5 are known apart
from the small Leslie coefficient $\alpha_1$.
$\alpha_1$ and some remaining uncertainties in the equilibrium  
conductivity
are fixed by fitting $\Vrms_c$ and $\v{q}_c$ within the SM.
Under the assumption
of a small recombination rate (as in I52), this enables a
quantitative test of the dependence
of $\wh$
on $\wo$.
The three layer thicknesses allow a test of the
predicted $d$ dependence which is independent of the
material parameters.
Only the overall magnitude of $\wh$ is fitted once with
$(\mupp\mupm)^{1/2}$ entering as a multiplicative factor
into the WEM expression.
The predicted and the measured
Hopf frequency agree, within the experimental errors,
for all applied
frequencies and all three cells, while the magnitude of $\wh$
varies by a factor of more than 50 (see Fig. 4).

\medskip

In Section II, we describe the measurements,
and in Section III, the relevant predictions of the WEM.
Theory and experiment are compared in Section IV.
Section V concludes with a discussion and points at some observed
nonlinear effects
which may be contained in the WEM.

\section{The Experiment}

We used three cells labelled C, D, and
E, respectively with areas of the
rectangular plate electrodes between
$10 \times 14$mm$^2$ and
$14 \times 16$mm$^2$ defining the directions $x$ and $y$.
Planar alignment along $x$ was obtained with a rubbed polyimide film
which coated the transparent electrodes.
The cell thickness in the $z$ direction
was measured on the empty cells by an infrared spectrophotometer.
Due to the large spotsize of the illuminating light beam this method
yields an average thickness of
$\overline{d}_C$ = 24.3 \mum, $\overline{d}_D$ = 14.7\mum, and
$\overline{d}_E$ = 29.3 \mum .
Interference fringes indicated that the local thickness varied by
$\pm2$ \mum\  around these values throughout the cell due to
deformation of
the plates at sealing.

The capacitance and conductance of the cells were measured by a  
capacitance
bridge shortly after filling with the nematic mixture Phase 5  
(Merck) as well
as after finishing the experiments (6 weeks later). Comparing these  data
with the
capacitance of the empty cells and neglecting stray capacitances,
the dielectric
permittivity $\epsp$ and the electric conductivity $\sigp\eq$
at equilibrium could be obtained.
These data are summarized in Table \tabcondexp.

The cells prepared for the measurements are too thin to determine
the exact values of the anisotropies of the dielectric permittivity
and the electric conductivity. Therefore we measured these parameters 
in a magnetic field using a thick ($>400$ \mum) cell calibrated  
with benzene.
Table \tabcondthick\ shows the resulting data
as a function of temperature.

The experimental apparatus for studying travelling waves consisted of
a temperature-control stage with a stability of $0.01$ K,
electronics for applying the ac voltage, and
a video camera with a computer-controlled imaging system.
All measurements were made at $(30\pm 0.01)\cels$.

The convection rolls were visualized by
the shadowgraph technique \cite{rasenat-shadow}.
The cell was illuminated by polarized
light with polarization along the director,
and the resulting shadowgraph signal
was monitored by a video camera
attached to a microscope using a 10x objective.
The images  correspond to an area of $257\mum \times 271 \mum$ of the cell.
The microscope was focused to the top of the cell
to allow for a dominant linear contribution of the signal
\cite{rasenat-shadow},
where the wavelength of the shadowgraph signal
coincides with that of the
director distortion. Second harmonics due to the nonlinearities
of the imaging technique could
be seen well above threshold.
The video camera was positioned to get an image
of vertical rolls in the normal-roll regime where $q_y=0$,
i.e., the rows of the
pixels were parallel to the wave vector of the pattern
in this regime.
A few measurements were carried out in the regime of  
oblique rolls, so
the ac frequency, where the transition between normal and oblique rolls
occurs (Lifshitz point),
could be estimated.

In the experimental runs, shadowgraph images were recorded
for each cell at various ac frequencies
in the conductive regime.
The measuring program for one frequency was as follows.
At first, the threshold voltage $\Vrms_c$
was roughly located by an automatic iterative search.
Then, a background image was recorded much below threshold.
This background was
subtracted on-line from all later images to remove
intensity inhomogeneities
due to illumination and surface irregularities.
To obtain an improved value of the threshold voltage,
the voltages were scanned in steps of $\Delta \eps \approx 0.002$
from a value above threshold $(\eps \approx 0.02)$ to a value below
$(\eps \approx  -0.02)$, and vice versa.
Here, the control parameter $\eps = \Vrms^2/\Vrms_c^2 - 1$ denotes the
deviation from the threshold.
Image recording followed the change of the applied voltage with a
delay of about 60-90 seconds.

The wavenumber in the x direction $q_x$
and the frequency
of the TWs at a given value of $\eps$ were determined
by Fourier transforms of a space-time image.
To obtain the space-time image,
a particular row was selected in the central portion of
a series of 256 images taken at consecutive time steps
and this row was copied from all these images into consecutive
rows of a new image yielding a 256 by 256 space-time plot.
Figure 1 shows such a plot for three values of $\eps$.
Figure 1a indicates that small-amplitude fluctuating patterns
could be observed slightly below threshold.
Travelling waves appear in these images as stripes tilted
toward the direction of travelling.
Both left and right travelling directions can be located. 
The horizontal periodicity
in these space-time images equals to $2 \pi/q_x$,
while the vertical one corresponds
to the period $1/f$ of the travelling waves. 

To increase the resolution, the Fourier
transformation was carried out on 2048
points with the first 256 points taken from the image (after
subtracting the average intensity), and the remaining points padded with 
zeroes.
For each row, the power spectrum $P(q_x)$ of the spatial frequencies
(i.e., the magnitude
of the squared Fourier amplitude)
was calculated and averaged over all rows. The wavenumber
$\overline{q}_x$ of the pattern was determined by a weighted
average  over all
points in the neighbourhood of the peaks of the power spectrum
corresponding to the
first and second spatial harmonics,
\begin{equation}\label{expk}
\overline{q}_x =
  \frac{\Sigma' q_x \overline{P}(q_x) +
        \frac{1}{2} \Sigma" q_x \overline{P}(q_x) }
       {\Sigma' \overline{P}(q_x) +
         \Sigma" \overline{P}(q_x)  }.
\end{equation}
Here, $\overline{P}(q_x)$ is the average of the power spectrum over all
rows attributed to the wavenumber $q_x$, $\Sigma'$ and $\Sigma"$ sums 
over all points with a maximum distance of
eight points from the peaks of the first and second harmonics,  
respectively.
The Hopf frequency is determined by analogous transformations and  
averaging
procedures for the columns.

The denominator of Eq. \refkl{expk},
\begin{equation}\label{amplit}
  \Sigma' \overline{P}(q_x) +
         \Sigma" \overline{P}(q_x)
  :=  \overline{A}^2,
\end{equation}
can be regarded as a measure of the averaged modulation intensity
of the shadowgraph signal.

For small distortions, the modulation amplitude
$\overline{A}$
is proportional to the amplitude of the director distortions
\cite{rasenat-shadow}.
The amplitude of the director distortions
is expected to scale with $\eps^{1/2}$ for
continuous bifurcations, at least, if
subcritical fluctuations play no role.
For $\eps$ large enough to neglect subcritical
fluctuations, but small enough to remain within the linear regime
of the shadowgraph technique,
one has $\overline{A}^2 \propto \eps$.
Figure 2 shows a typical example of $\overline{A}^2$ vs. $\eps$
for cell D at $f_0=$ 90 Hz corresponding
to $\wo\tau_q = 2.06$.
The threshold was defined by extrapolating the first roughly
linear rise of $\overline{A}^2$
to zero (dashed line).
For $\eps > 0.09$ one enters the nonlinear regime 
of the shadowgraph, which is indicated by the
appearance of a second harmonic peak in the Fourier spectrum.
The scatter in the data of Fig. 2 for $\eps > 0.011$ is a result
of the disorder appearing in the travelling pattern.
There are structureless (nonperiodic) areas separating the left and
right travelling waves with random space and time distribution which reduce
the averaged intensity modulation accidentally.

The wavenumbers and the frequencies were calculated
according to Eq.(1) as a function of $\eps$.
The values $q_{cx}$ and $\wh$ at threshold ($\eps=0$)  are obtained by
fitting a straight line
to points in the range $-0.01<\epsilon<0.02$.
To monitor the change of the conductivity during the measurements,
the transition frequency
$\omega\boxsub{cd}$ to dielectric rolls (see Sec. IV below)
was determined at the start and at the end of the experimental run
for each cell. $\omega\boxsub{cd}$ at the time of the measurement
was determined from these values by a linear interpolation.

Despite careful preparation of the cells,
there was a spatial variation of the thresholds, i.e., on
increasing $\Vrms$ from subcritical values,
convection did not fill the whole cell simultaneously.
To obtain the most homogeneous image possible,  a local minimum or  
maximum
of the threshold was selected in the cell
for the monitored area of the measurements.
In cell C, a local minimum was chosen (convection sets in first),
while in the cells D and E,
a local maximum was chosen (convection sets in last;
rolls appeared
first at the edges of the electrode).

\medskip

Elastic moduli of the substance were determined by measuring the
Freedericksz thresholds in various geometries
\cite{degennes,deuling}.
The magnetic field induced splay and twist Freedericksz transitions
could be observed on the same planar cells used for
studying the travelling waves.
For magnetic and electric field induced bend transitions
homeotropic cells were prepared.
The field induced deformation was detected by optical methods
(birefringence for splay and bend, depolarization for twist transitions).
As the anisotropy of the magnetic susceptibility is unknown for Phase 5
the measurements of the threshold magnetic field yielded only a ratio
of the elastic moduli,
$\frac{K_{22}}{K_{11}} = 0.47$ ;
$\frac{K_{33}}{K_{11}} = 1.29$.
Measuring the threshold voltage in homeotropic cells we obtained
$K_{33} =12.7 \times 10^{-12} N$.
The error of these measurements, which are summarized in Table  
\tabmatpar,
is $\pm10$\%.

%
{\tabsetting
\begin{table}
\caption{Conductivity, dielectric permittivity, and related quantities
in the cells used for the EHC experiments ($T = 30 \cels$).}

\vspace{3mm}

\begin{tabular}{c|c|c|c|cc}  
Quantity
  & Cell C
  & Cell D
  & Cell E
  & Comments
  & \spacingmid \\ \hline
Average thickness $\overline{d}$ $[\mum ]$
  & 24.3
  & 14.7
  & 29.3
  & There is a $\pm2 \mum $ deviation
  & \spacingtop \\ 
  &
  &
  &
  & from $\overline{d}$ throughout the cell
  & \spacingbot \\ \hline
Local thickness $d$ $[\mum ]$
  & 23.5
  & 13.4
  & 27.5
  & Deduced from the wavevector (see Sec. IV)
  & \spacingmid \\ \hline
Conductivity $\sigp\eq$
  & 8.5
  &16.4
  & 7.3
  & in fresh cell
  & \spacingtop \\
$[10^{-9}\Ommi]$
  & 8.6
  &14.7
  &11.1
  & 6 weeks later
  & \spacingbot \\ \hline
Permittivity $\epsp$
  & 5.34
  & 5.38
  & 5.24
  & in fresh cell
  & \spacingtop \\
  & 5.30
  & 5.39
  & 5.19
  & 6 weeks later
  & \spacingbot \\ \hline
Range of the transition
  & $84 - 90$
  & $138 - 177$
  & $112 - 130$
  & minimum and maximum values
  & \spacingtop \\
frequency $\omega\boxsub{cd}$ [Hz]
  &
  &
  &
  & during measurements
  & \spacingbot \\ \hline
$\overline{\sigma}$ $[10^{-9}\Ommi]$
  & 7.0   
  & 13.0
  & 8.0
  & Average conductivity obtained from
  & \spacingtop \\
  &&&
  & the measured $\omega\boxsub{cd}$ (see Sec. IV)
  & \spacingbot \\
\end{tabular}
\end{table}
} 
%
%

%
{\tabsetting
\begin{table}

\caption{Dielectric permittivities  and electric conductivities
of Phase 5 as a function of temperature for a thick
cell ($400 \mum$) at 1 kHz}

\vspace{3mm}

\begin{tabular}{ccccccccl}
$T (\cels)$
    & $\epsp$
    & $\epspar$
    & $\ea$
    & $\sigp\eq (10^{-9} \Ommi)$
    & $\sigpar\eq (10^{-9} \Ommi)$
    & $\siga\eq (10^{-9} \Ommi)$
    & $\siga\eq/\sigp\eq$
    & \spacingmid \\ \hline
 20
    & 5.336
    & 5.100
    & -0.236
    & 2.89
    & 4.98
    & 2.09
    & 0.723
    & \spacingtop \\
 22
    & 5.319
    & 5.086
    & -0.233
    & 3.07
    & 5.35
    & 2.28
    & 0.743
    & \\
 23
    & 5.305
    & 5.084
    & -0.221
    & 3.29
    & 5.65
    & 2.36
    & 0.717
    & \\
 25
    & 5.286
    & 5.079
    & -0.207
    & 3.66
    & 6.44
    & 2.78
    & 0.760
    & \\
 27
    & 5.256
    & 5.058
    & -0.198
    & 4.25
    & 7.38
    & 3.13
    & 0.736
    & \\
 30
    & 5.217
    & 5.033
    & -0.184
    & 5.16
    & 8.73
    & 3.57
    & 0.692
    & \\
 32
    & 5.185
    & 5.013
    & -0.172
    & 5.89
    & 10.17
    & 4.28
    & 0.727
    & \\
 35
    & 5.159
    & 4.994
    & -0.165
    & 6.69
    & 11.22
    & 4.53
    & 0.677
    &
\end{tabular}
\end{table}
} 




\begin{figure}

\figps{140mm}{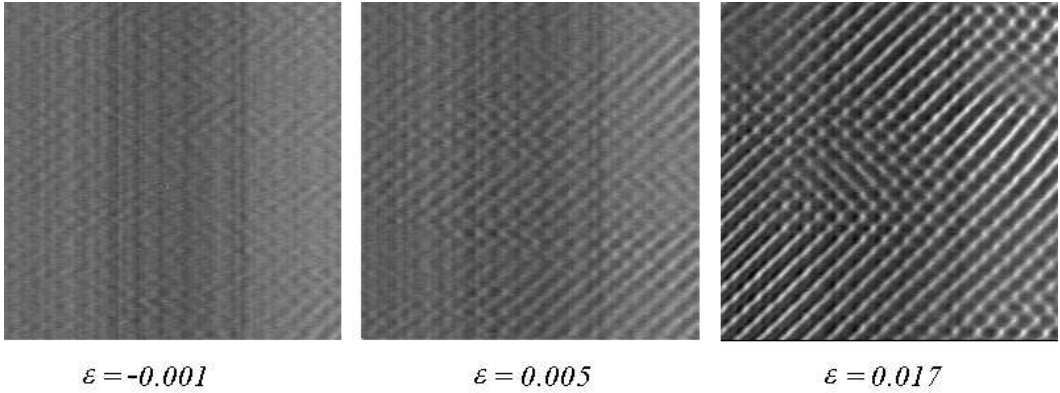}

\vspace{3mm}

\caption{Space-time images of the travelling waves in cell D at  
$f=90$ Hz.
The contrast was enhanced by a factor of 5.3 for all three images.
(a) Below threshold ($\eps=-0.001$). Subcritical fluctuations can  
be seen.
(b) Sligthly above threshold ($\eps=0.005$).
(c) Further above threshold ($\eps=0.017$). Travelling
waves in both directions are
observable.
}
\end{figure}


\begin{figure}

\figps{120mm}{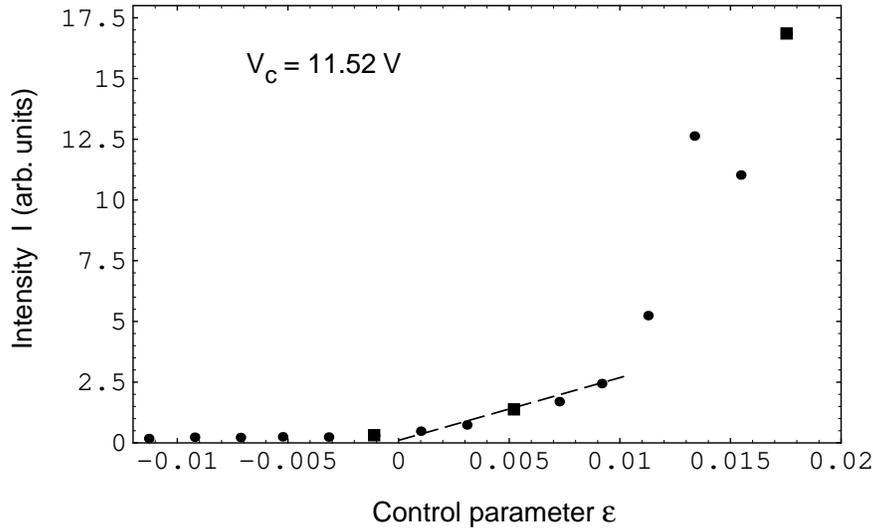}

\vspace{3mm}

\caption{Averaged modulation intensity $ I = \protect\overline{A}^2$ of 
the shadowgraph signal as a
function of applied voltage for cell D at $f=90$ Hz.
The threshold $\protect\Vrms_c$ is defined by extrapolating the
first roughly linear rise of $I$ to zero (dashed line).
The three squares denote the points corresponding to
the space-time images in Fig. 1.
}
\end{figure}

   \section{Predictions of the weak electrolyte model}

In the WEM, the conductivity of the NLC is described by the drift
of two species of oppositely charged freely mobile ions
originating from a dissociation-recombination reaction
from neutral molecules.
The ionizable molecules are not specified.
They can be
(uncontrolled) impurities (which is assumed in most experiments on
MBBA and Phase 5 and presumably accounts for the strong variations of the
conductivity),
or ionizable dopants
(which is more controlled and realized, e.g.,  in I52
\cite{dennin-treiber-PRL,dennin-ILCC}).

The WEM assumes that 
only a small fraction of the impurities is dissociated into
ionic charge carriers (weak electrolyte).
This is not an essential restriction but it allows to
reduce the number of material parameters not contained in the SM.
For typical values for the
mobilities [$10^{-10}\miiVs$] and conductivities
$[10^{-8}\Ommi]$, the number density of both species of
ions is about $3 \times 10^{20} \mbox{m}^{-3}$.
For a molecular weight
of the impurities of 100, this corresponds to a concentration 
of 0.05 ppm.
One has a weak electrolyte, if the number density of the
impurity molecules is sufficiently higher. 
In the experiments employing I52 mentioned above, this is
certainly fulfilled \cite{treiber-PhD}.
In the present work, the weak electrolyte assumption
is consistent with the roughly exponential increase of
the conductivities with temperature (Table II) corresponding
to a dissociation energy of about 0.4 eV.

The drift velocities of the oppositely charged ions (superscript $+$  
and $-$)
relative to the fluid under the influence
of the electric field $\v{E}$
are given by
\begin{equation}
\label{mu}
\v{v}\boxsub{drift}^{\pm} = \pm \m{\mu}^{\pm} \v{E}, \ \ \ \
     \mu_{ij}^{\pm} = \mu_{\perp}^{\pm}
     \left(\delta_{ij} + \sigma'_a n_i n_j \right),
\end{equation}
defining the uniaxially
anisotropic mobility tensors for the two ions.
Here, $\delta_{ij}$ is the Kronecker symbol,
$n_i$ and $n_j$ are components of
the director,
and $\sigma'_a = \sigma_a\eq/\sigp\eq$ is the relative
anisotropy of the conductivity.
The local conductivity perpendicular to the director
(which is constant in the SM), is given in terms of the
number densities $n^{\pm}$.
For charges $\pm e$, one obtains
\begin{equation}
\label{sigp}
\sigp\vonrt = e \left( \mupp n^+ + \mupm n^- \right).
\end{equation}
The basic equations of the WEM are given in Ref. \cite{treiber-ILCC}.
They are obtained by transforming the rate equations
for the ionic number densities $\np$ and $\nm$ into
an equation for $\sigp\vonrt$ describing the dynamics
of the drift and of the dissociation-recombination reaction,
and a conservation equation for the charge density
$\rho = e(\np+\nm)$ together
with the Poisson equation.
The balance equations for the director components and
the generalized Navier--Stokes equations for the
fluid velocities are those of the SM.

The non-convecting basic state is nearly homogeneous, which  
simplifies the
linear stability calculations.
A one-mode Galerkin approximation for the linearized fields
yields quantitative results which typically deviate by less than
3\% from the fully numerical solution \cite{treiber-PhD}.
In this approximation, the dependence
of each field on $z$ and $t$ is represented by one test function
satisfying the boundary conditions and the symmetries under reflection
and time translation applying
in the conductive
regime \cite{treiber-ILCC}.
This one-mode approximation enables analytic formulas for the
Hopf frequency, and for the
threshold shift $\Delta\eps \equiv (\Vrms_c/\Vrms_c\SM)^2-1$
of the WEM with respect to the SM.
For details in the normal-roll regime, see Ref. \cite{treiber-ILCC};
the generalization to oblique rolls can be found in
Ref. \cite{dennin-treiber-PRL}.
The threshold shift is always smaller than 1\% for the experiments
of this paper, so the thresholds, and also the critical wavevectors,
are calculated using the SM.

For $\wschl \tau_d << \Vrms^2 \epso \ep / (K_{11}\pi^2)$
which is fulfilled for all measurements,
the analytic expression for
$\wh = 2 \pi f_H$ is
\begin{eqnarray}
\label{wh}
  \omega_H  &=& \wschl
  \sqrt{1 - \frac{1}{(\tau_{rec} \wschl)^2}}, \\
\label{wschl}
 \wschl  &=&
       \pi C' \frac{\Vrms_c^2 \epso\epsp}{d^3(1+\ws^2)}
       \sqrt{\frac{\mu_{\perp}^{+} \mu_{\perp}^{-}}{\gamma_1\sigp\eq}}.
\end{eqnarray}
Here,
$C' = \sqrt{\siga'} C $ is a dimensionless factor of order unity
with $C$ given for normal rolls in Ref. \cite{treiber-ILCC} (see also
\cite{treiber-ILCCerr}),
and $\ws= \wo\tau_q
  (1+ \v{q}_c^{'2} + \eps_a'   q_{cx}^{'2}) /
  (1+ \v{q}_c^{'2} + \sigma_a' q_{cx}^{'2})$
with $\epsa' = \epsa/\epsp$, $\v{q}=(q_x,q_y)$, and
$\v{q}' = \v{q} d / \pi$.
In the experiments described here, one has always
$0.45 < C' < 0.75 $ and
$0.60\ \wo\tau_q < \ws < 0.68\ \wo\tau_q $.

%



The main ingredients of Eqs. \refkl{wh} and \refkl{wschl}
can be seen by considering the two feedback mechanisms
resulting from the WEM.
The primary instability
mechanism is the Carr--Helfrich mechanism of the SM.
Under an applied field, an initial director bend leads \via\ the anisotropy
of the conductivity to a charge
accumulation and to an electric volume force driving the fluid motion.
The resulting velocity gradients couple back to the director
\via\ the rotational viscosities.
The time scale of this mechanism is $\sim \tau_d$.
The gradients of the charge accumulation, however,
also excite the charge-carrier mode whose feedback tends to decrease
the charge.
The time scale of this second stabilizing drift mechanism is
$\sim \tau_t^2/\tau_q$.
As in binary mixtures (see, e.g., \cite{cross-hohenberg}), the
interplay of the two mechanisms can lead to oscillations with a (Hopf)
frequency proportional to the geometric mean of the two
processes involved. This gives
$\omega \sim \sqrt{\tau_q/(\tau_d\tau_t^2)}$, which is Eq. \refkl{wschl},
up to a dimensionless prefactor.

The drift mechanism is only effective if the charge-carrier mode
can build up sufficiently, i.e., for a small recombination rate.
This leads to the square root in Eq. \refkl{wh} and to the
condition $\tau\rec^{-1} < \wschl$ for the occurrence of a Hopf
bifurcation to TWs.
%
%
The condition $\wschl = \tau\rec^{-1}$
gives the codimension-two point separating
the regime of stationary rolls ($\tau\rec^{-1} > \wschl$) from that of TWs.
This condition scales with $d^{-3}$, $(\sigp\eq)^{-1/2}$, and
$\Vrms_c^2/(1+\ws^2)$, hence a Hopf bifurcation is favoured by
thin cells, clean cells (low $\sigp\eq$), and 
by high values of
the applied frequencies in case of materials with $\ea < 0$
(for $\ea < 0$, the factor $\Vrms_c^2/(1+\ws^2)$ increases
with $\wo$
and contains the main part of the dependence on $\wo$).

In contrast to former experiments on Phase 5
\cite{rehberg-sol,ribotta-88},
we observed
TWs for all applied frequencies
in the three cells. This is presumably due to a smaller concentration
of impurities.
The Hopf condition is most restrictive for the thickest
cell at the lowest frequencies.
The experimental points for this cell
(squares in Fig. 3a)
suggest that one is still far
away from the codimension-two point.
With Eq. \refkl{wh}, this means that $\wh\approx\wschl$, i.e.,
$1 / \tau\rec$, can be set to zero in the calculations.

In this case,
the WEM predictions contain $(\mupp\mupm)^{1/2}$
as the only non-SM parameter,
which appears  in $\wh \approx \wschl$,
as a simple multiplicative factor.
The WEM predicts that $\wschl$
is proportional to $d^{-3} (\sigp\eq)^{-1/2}$ and increases
with the ac frequency.
In the following section,
these predictions will be compared
with the experiment.

\section{Comparison between experiment and theory}

The WEM gives $\Vrms_c$, $\v{q}_c$, and the Hopf frequency $\wh$,
as a function
of the applied frequency $\wo\tau_q$
normalized to the inverse of the
charge relaxation time $\tau_q$.
The values for $\Vrms_c$ and $\v{q}_c$ differ by less than 1\% from
those of the SM.
In a good approximation, $\Vrms_c(\wo\tau_q)$ and
$q_c(\wo\tau_q) d/\pi$ do not depend on $d$ or
on $\sigp\eq$ \cite{kramer-jphys}, the only parameters
that are significantly different in the three cells.
All experimental points for $\Vrms_c$ and $q_c$
should lie on one curve, if one plots $\Vrms_c$
and $q_c d/\pi$ against $\wo\tau_q$.
Similarly, the WEM requires that the experimental
points for $\wh d^3 (\sigp\eq)^{1/2}$ should lie on one line.
To verify this,
requires knowledge of $\tau_q \propto 1/\sigp\eq$
for each experimental point.

The conductivity turned out to vary somewhat irregularly
during the experimental runs
(Table \tabcondexp).
The relative change of the conductivities
during the measurements can be assessed by determining,
for each measurement of the set $\Vrms_c$, $q_c$, and $\wh$,
the transition
frequency $\omega\boxsub{cd}$ where the rolls of the  conductive
regime cross over to dielectric rolls \cite{kramer-EHCrevs},  which  
can be
easily identified.
This frequency is related to the "cutoff frequency"
$\omega\boxsub{cutoff}$ where the threshold of the conductive  
regime diverges
in the one-mode approximation.
The ratio $\omega\boxsub{cd}/\omega\boxsub{cutoff}$ depends on $d$,
but is constant for one cell
($\omega\boxsub{cutoff}/\omega\boxsub{cd} = 0.68 \cdots 0.74$
for the three cells).
Since the cutoff frequency in the SM
is proportional to $\sigp\eq$
($\omega\boxsub{cutoff} = 5.57 \tau_q^{-1}$ for the parameters of
Table \tabmatpar),
the actual conductivity can be written as
\begin{equation}\label{sigpex}
\sigp\eq = \frac{\omega\boxsub{cd}} {\overline{\omega}\boxsub{cd} }  
\ \ \barsig,
\end{equation}
where $\overline{\omega}\boxsub{cd}$ is defined as
the average of the transition frequencies
over all measurements in one cell (C,D, or E)
and the "average conductivity" $\barsig$ must be determined
separately.

Figure 3a shows the resulting threshold voltages
(from the SM) as a function
of $\wo\tau_q = \wo \epso\epsp/\sigp\eq$
with $\sigp\eq$ from Eq. \refkl{sigpex} (continuous curve) together with
the experimental results.
For each of the cells, $\barsig$ was
fitted  to the threshold curves of the SM
with the other parameters taken from Table \tabmatpar\ in Appendix A.
Of course, the resulting value of $\barsig$
must be consistent with the
independent conductivity measurements
(Table \tabcondexp) taken some time before and after
the measurements of the patterns.
Table \tabcondexp\ shows that
$\overline{\sigma}$ deviates by
less than 25 \% from these measurements, which is within the
variations of the conductivities as estimated from the variations of
$\omega\boxsub{cd}$.
The unknown viscosity $\alpha_1$ was adjusted as well.
In the normal-roll regime (all but the measurements
corresponding to the three leftmost triangles are in this regime),
the influence of $\alpha_1$ on the threshold
and on $\wh$ is negligible.
In contrast, the tendency to oblique rolls is increased by increasing
$\alpha_1$; so $\alpha_1$ was fitted to give the correct frequency for 
the transition from normal to oblique rolls, the Lifshitz point (in  
the SM).

Figure 3b shows the normalized $x$-component $q_{cx} d/\pi$ of the
wavevector.
Taking for $d$ the average thicknesses measured on the empty cells
($\overline{d}$,  Sec. II) gives
a systematic deviation
(5\% - 10 \%) of the experimental points from the
curves from the SM. One can attribute this to the fact that
the actual thickness at the position of the measurement may differ from
$\overline{d}$ (see Sec. II). 
The systematic deviations can mostly be removed by using 
slightly reduced thickness values
$d_C$ = 23.5 \mum, $d_D$ = 13.4\mum, and  $d_E$ = 27.5 \mum
(Table \tabcondexp).
This correction (a reduction by  0.8 \mum\ for cell C,
by 1.4 \mum\ for cell D, and 1.8 \mum\ for cell E) is within
the limits of deviation indicated by the interference fringes ($\pm2$ \mum).
Hence for the plots presented in this paper we assumed
these reduced thickness values.
The agreement for these thicknesses is reasonable, although the
theoretical frequency dependence (from the SM) is somewhat
too steep at larger $\wo\tau_q$.
The kink in the theoretical curve
at $\wo\tau_q = 0.8$ is due to the Lifshitz point.
%

In Figure 4, we show the main result of this paper,
namely  the measured
Hopf frequencies together with the WEM predictions,
Eq. \refkl{wh}, under the assumption of a long
recombination time where $\wh = \wschl$.
The overall magnitude of $\wh$ fixes the mobilities to
\begin{equation}\label{muphase5}
\geomu = 1.1 \times 10^{-10} \miiVs,
\end{equation}
but nothing else can be fitted.
Since in the normal-roll regime $\alpha_1$ plays
a minor role for the dependence on $\wo$,
the $d$ dependence and the dependence on $\wo$
provided by the theory are in effect
free from unknown or adjustable
parameters.

The measurements 
for the thickest cell (squares) and the thinnest cell (triangles) 
agree
with the WEM curve nearly everywhere within the experimental error of
about 25\%; the Hopf frequency of cell C (circles) is systematically
higher ($\approx 25\%$) than the predicted one.
This has to be compared with the wide range covered by 
the actual Hopf frequencies.
In the experiments, the Hopf frequency varies from
about $0.4 \mbox{s}^{-1}$ (cell E at $\wo\tau_q=0.8$) to more than
$22 \mbox{s}^{-1}$ (cell D at $\wo\tau_q=4.0$),
i.e., by a factor of more than 50.
For a given external frequency, the Hopf frequency of cell D is
about seven times that of cell E.

%
\newpage


\begin{figure}
\figps{140mm}{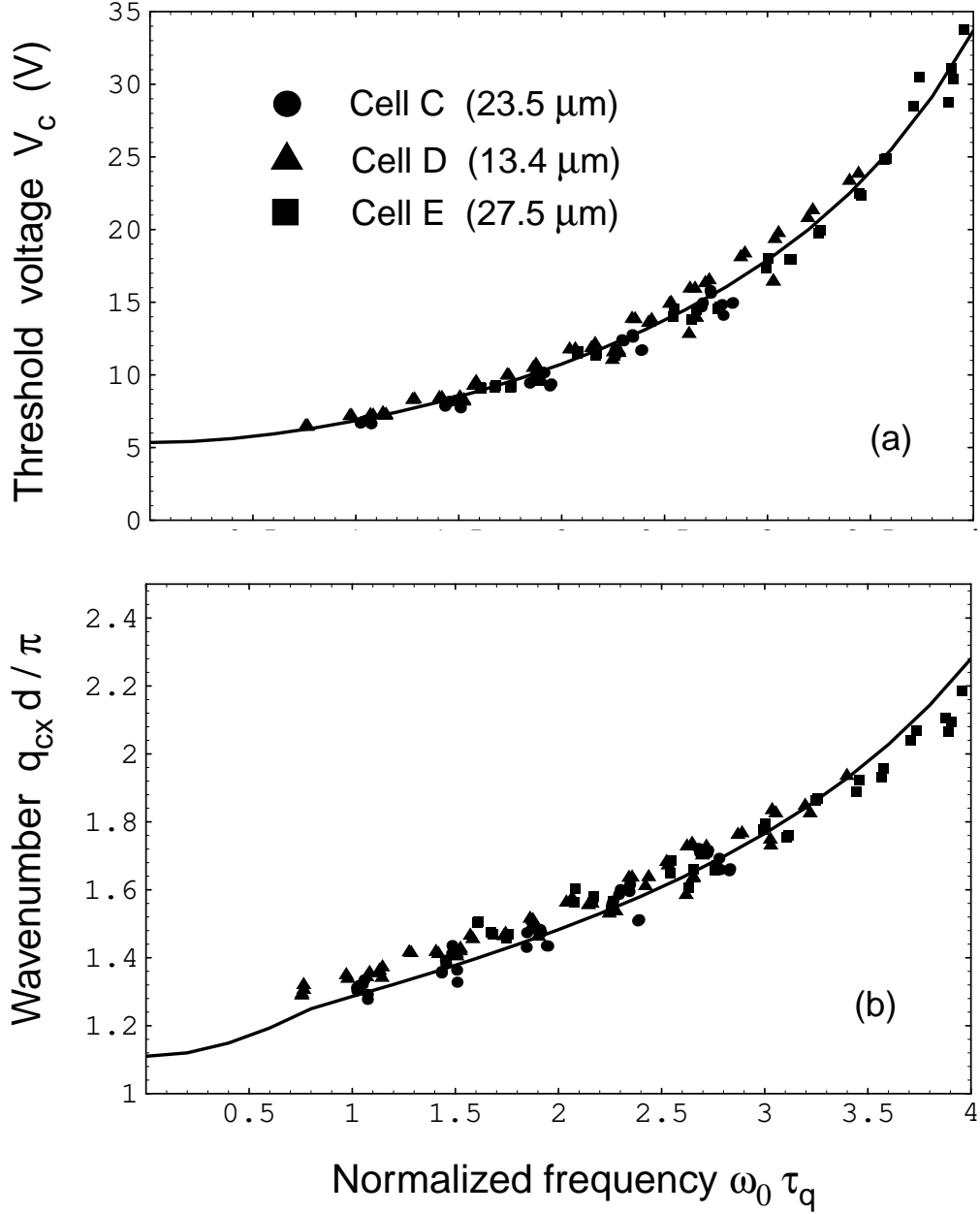}

\vspace{3mm}

\caption{(a) Threshold voltage,
and (b) wavenumber $q_{cx}$
as a function of
$\protect\omega_0\protect\tau_q$.
The conductivity $\protect\sigp\protect\eq$ in $\protect\tau_q$
is calculated individually for each point
from $\protect\omega\protect\boxsub{cd}$,
see the main text.
The experimental points of the three
cells (circles, squares, and triangles) should lie on one curve
for $\protect\Vrms_c$ and $q_{cx} d/\pi$, respectively.
The curves are the SM predictions for the material parameters of Phase 5
given in Appendix A with the viscosity $\protect\alpha_1$
fitted to $-0.35 \protect\gamma_1$.
The WEM predictions for $\protect\Vrms_c$ and
$q_{cx}$ cannot be distinguished from
that of the SM on this scale.
}
\end{figure}

\begin{figure}
\figps{150mm}{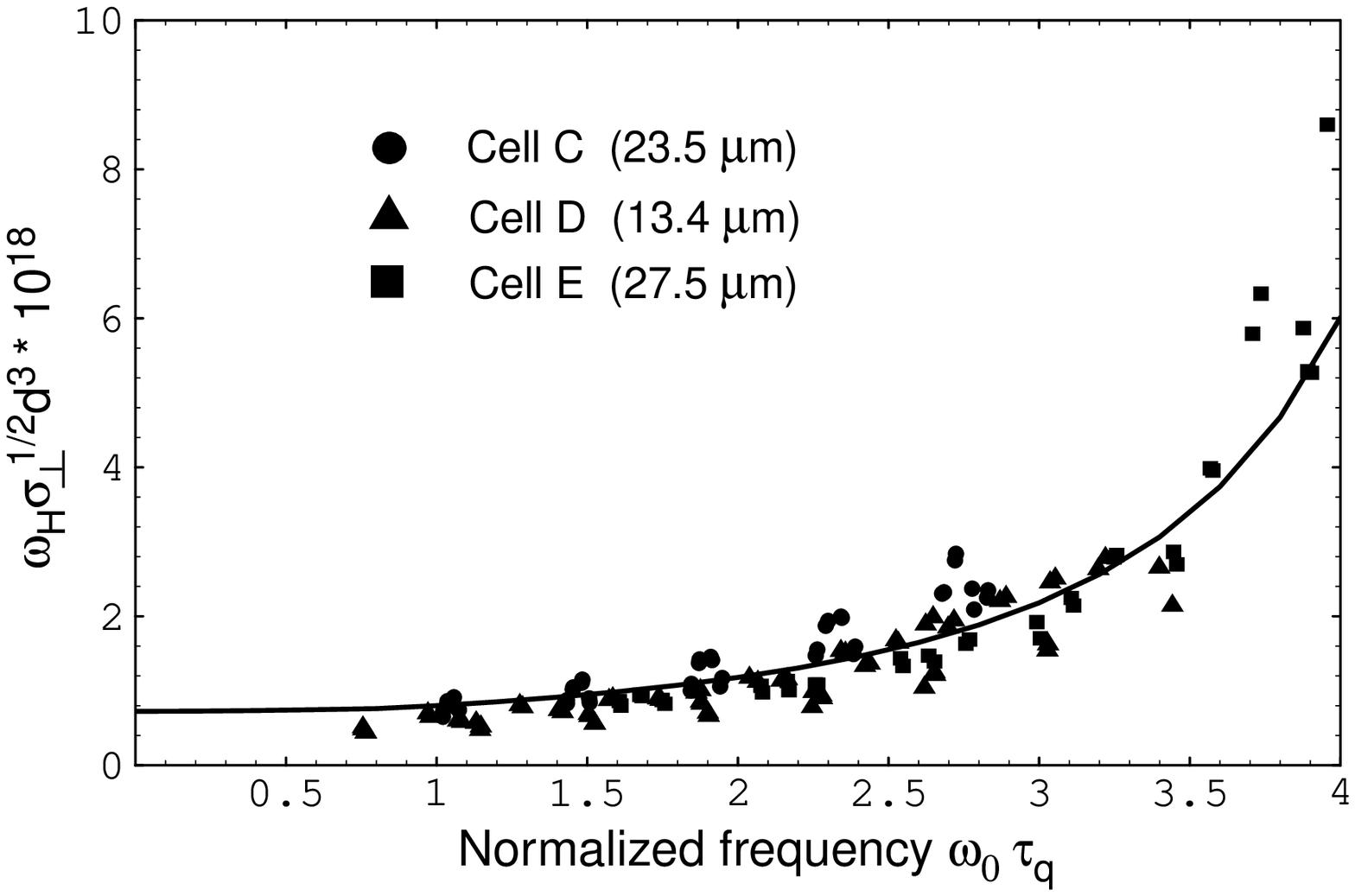}

\vspace{3mm}

\caption{The Hopf frequency of the travelling rolls multiplied by
$(\protect\sigp\protect\eq)^{1/2} d^3$, is shown as a function of
the normalized applied frequency $\protect\wo\protect\tau_q$.
The circles, squares, and triangles denote the experimental
results for the three cells. The curve is the universal
WEM prediction
for the material parameters of Phase 5 (Appendix A)
and $\protect\geomu = 1.1 \times 10^{-10} \protect \miiVs$.
For $\protect\wo\protect\tau_q = 0$, this curve corresponds to
$\protect\wh =$
0.67   s$^{-1}$,
2.65   s$^{-1}$, and
0.37   s$^{-1}$,
for the cells C, D, and E, respectively.
}
\end{figure}

\section{Conclusion}

Our results represent the most stringent test for the  
WEM model so far.
The frequency dependence of the threshold $\Vrms_c$ and,
to less extent, the  
critical wavevector
$\v{q}_c$, involving in essence only SM quantities,
has been tested previously fairly well in the conductive range
\cite{kramer-EHCrevs,kramer-EHC89}.
Also $\alpha_1$ was fitted to obtain the correct Lifshitz point, but this
is of little influence in the normal-roll range considered here mainly.
The Hopf frequency $\wh$ predicted by the WEM deviates from
the measured ones by at most $\approx$ 25\% in a range
of parameters where $\wh$ varies by a factor of about 50. This was  
achieved
by fitting only the parameter $\geomu$ appearing as an overall factor.

The value of $\geomu$ given in Eq.(\ref{muphase5}) compares  
reasonably well
with a number of measurements of the quantity $\mu = \mupp+\mupm$ in different 
materials, namely (in units of $10^{-10}\miiVs$)
$\mu$
= 0.18 \cite{briere}, 1 \cite{blinov-old},
3.71 \cite{richardson}, 1..10 \cite{degennes-sol}, 10 \cite{turnbull}. 
Little appears to be known about the ratio $\mupp/\mupm$.

Our results indicate that the recombination time $\tau\rec$
is well above $1/\wh$ even for the smallest Hopf frequency measured,
which means $\tau\rec$ is larger than about $5$s. Such long times
may seem surprizing, however, literature values for MBBA range from
$10^{-3}$s \cite{turnbull} to $2.7 \times 10^4$s \cite{briere}.
Thus in our experiments one is well away from the codimension-2
point that separates the stationary regime from the Hopf regime.
Let us point out that a weakly nonlinear analysis
predicts the bifurcation to be typically subcritical
near the codimension-2 point on the stationary side 
\cite{treiber-PhD}. 
This appears
to explain the slight hysteresis observed in some high-sensitivity
experiments 
\cite{dennin-treiber-PRL,rehberg-PRA-91-fluct,rehberg-PRL-91-fluct}.
The fact that a hysteresis is sometimes
also observed on the Hopf bifurcation side 
\cite{rehberg-PRA-91-fluct,rehberg-PRL-91-fluct} is not  
disturbing
because there is evidence that this is due to higher-order effects
in the amplitude \cite{hoerner-priv}. The theoretical analysis
is consistent with this fact.

For further testing of the WEM
an independent determination of both $\tau\rec$ and the mobilities
is desirable. This could be done
by measuring the transient current response to suitable voltage
signals in cells with well defined boundary conditions
(e.g., blocking electrodes) \cite{naito-91,sugimura,briere}.

\begin{acknowledgements}

We wish to thank J. Peinke and M. Scheuring for help with the
experiments and
with the characterization of the material.
R. Stannarius made us aware of Ref.\cite{kneppe} where the
viscosities of Phase 5 were measured.
A.B. and N.\'E. appreciate the support and kind hospitality of the
University of Bayreuth.
The nematic Phase 5 was kindly made available for the measurements by Merck,
Darmstadt.
This work was financially supported by the Volkswagen-Stiftung,
the Hungarian Research Grant No. OTKA T014957
and the EU TMR project ERB FMRX-CT 96-0085.

\end{acknowledgements}

\newpage

\begin{appendix}
\section{Material parameters for Phase 5 at 30 \cels}
The parameters $K_{11}$, $K_{22}$, $K_{33}$, $\epsp$,
$\epsa$, $\sigp\eq$, and $\siga\eq$
were measured in this work.
The actual value of $\sigp\eq$ varied considerably and was determined
individually for each measurement by Eq. \refkl{sigpex} with
$\overline{\sigma}$ from Table \tabcondexp.
There are  published data for the dielectric anisotropy  
\cite{phase5-merck}
which differ somewhat, but these data seem to be
rather inprecise.
The Leslie viscosities $\alpha_2$ to $\alpha_6$ were
determined from the viscosities $\eta_1$, $\eta_2$, $\eta_3$, and
$\gamma_1$ of Ref. \cite{kneppe}
by the relations
$\alpha_2 = (\eta_2-\eta_1-\gamma_1)/2$,
$\alpha_3 = (\eta_2+\eta_1-\gamma_1)/2$,
$\alpha_4 = 2 \eta_3$,
$\alpha_6 = (\eta_1 + 3 \eta_2 - 4 \eta_3 - \gamma_1)/2$,
and the Onsager relation
$\alpha_5 = \alpha_6-\alpha_2-\alpha_3$.
%
%
\begin{table}

\caption{Material parameters for Phase 5}

\vspace{3mm}

\begin{tabular}{ccccl}
Parameter
    & Unit
    & Value at 30 \cels
    & Source and comments
    & \spacingmid \\ \hline
$K_{11}$
    & $10^{-12}$ N
    & 9.8
    & Sec. II
    & \spacingtop \\
$K_{22}$
    & "
    & 4.6
    & "
    & \\
$K_{33}$
    & "
    & 12.7
    & "
    & \spacingbot \\ \hline
$\siga/\sigp\eq$
    & --
    & 0.7
    & Table \tabcondthick
    & \spacingmid \\ \hline
$\epsp$
    & --
    & $5.19 \cdots 5.39$
    & Tables \tabcondexp, \tabcondthick
    & \\
$\epsa$
    & --
    & $ -0.184$
    & Table \tabcondthick; see also Ref. \cite{phase5-merck}
    & \spacingbot \\ \hline
$\alpha_1$
    & $ 10^{-3}$ Ns/m$^2$
    & $-39$
    & Fit to the Lifshitz point
    & \spacingtop \\
$\alpha_2$
    & "
    & $ -109.3$
    & Kneppe, et al, Ref. \cite{kneppe}
    & \\
$\alpha_3$
    & "
    & 1.5
    & "
    & \\
$\alpha_4$
    & "
    & 56.3
    & "
    & \\
$\alpha_5$
    & "
    & 82.9
    & "
    & \\
$\alpha_6$
    & "
    & $ - 24.9$
    & "
    & \spacingbot \\ \hline
$\sqrt{\mupp\mupm}$
    & $ 10^{-10} \miiVs$
    & 1.1
    & Fit to $\wh$
    & \spacingtop \\
$\tau\rec^{-1}$
    & s$^{-1}$
    & $<0.2$
    & Hopf bifurcation in all cells
\end{tabular}
\end{table}

\end{appendix}
%

\begin{thebibliography}{10}
\bibitem{treiber-ILCCerr}
There is a printing error in Eq. (38) of Ref. \cite{treiber-ILCC}. The factor
  $(1 - \frac{\epsilon_a}{\epsilon_q}L_{nn}q^2)$ should be replaced by $(L_{nn}
  - \frac{\epsilon_a}{\epsilon_q}q^2)$.
\end{thebibliography}


%
%
%

\end{document}